\documentclass[twoside,twocolumn,9pt]{article}
\usepackage{extsizes}
\usepackage[super,sort&compress,comma]{natbib} 
\usepackage[version=3]{mhchem}
\usepackage[left=1.5cm, right=1.5cm, top=1.785cm, bottom=2.0cm]{geometry}
\usepackage{balance}
\usepackage{times,mathptmx}
\usepackage{sectsty}
\usepackage{graphicx} 
\usepackage{lastpage}
\usepackage[format=plain,justification=raggedright,singlelinecheck=false,font={stretch=1.125,small,sf},labelfont=bf,labelsep=space]{caption}
\usepackage{float}
\usepackage{fancyhdr}
\usepackage{fnpos}
\usepackage[english]{babel}
\usepackage{array}
\usepackage{droidsans}
\usepackage{charter}
\usepackage[T1]{fontenc}
\usepackage[usenames,dvipsnames]{xcolor}
\usepackage{setspace}
\usepackage[compact]{titlesec}

\usepackage{epstopdf}

\definecolor{cream}{RGB}{222,217,201}

\begin{document}

\pagestyle{fancy}
\thispagestyle{plain}


\makeFNbottom
\makeatletter
\renewcommand\LARGE{\@setfontsize\LARGE{15pt}{17}}
\renewcommand\Large{\@setfontsize\Large{12pt}{14}}
\renewcommand\large{\@setfontsize\large{10pt}{12}}
\renewcommand\footnotesize{\@setfontsize\footnotesize{7pt}{10}}
\makeatother

\renewcommand{\thefootnote}{\fnsymbol{footnote}}
\renewcommand\footnoterule{\vspace*{1pt}%
\color{cream}\hrule width 3.5in height 0.4pt \color{black}\vspace*{5pt}} 
\setcounter{secnumdepth}{5}

\makeatletter 
\renewcommand\@biblabel[1]{#1}            
\renewcommand\@makefntext[1]%
{\noindent\makebox[0pt][r]{\@thefnmark\,}#1}
\makeatother 
\renewcommand{\figurename}{\small{Fig.}~}
\sectionfont{\sffamily\Large}
\subsectionfont{\normalsize}
\subsubsectionfont{\bf}
\setstretch{1.125} 
\setlength{\skip\footins}{0.8cm}
\setlength{\footnotesep}{0.25cm}
\setlength{\jot}{10pt}
\titlespacing*{\section}{0pt}{4pt}{4pt}
\titlespacing*{\subsection}{0pt}{15pt}{1pt}

\makeatletter 
\newlength{\figrulesep} 
\setlength{\figrulesep}{0.5\textfloatsep} 

\newcommand{\topfigrule}{\vspace*{-1pt}%
\noindent{\color{cream}\rule[-\figrulesep]{\columnwidth}{1.5pt}} }

\newcommand{\botfigrule}{\vspace*{-2pt}%
\noindent{\color{cream}\rule[\figrulesep]{\columnwidth}{1.5pt}} }

\newcommand{\dblfigrule}{\vspace*{-1pt}%
\noindent{\color{cream}\rule[-\figrulesep]{\textwidth}{1.5pt}} }

\makeatother

\twocolumn[
  \begin{@twocolumnfalse}
\vspace{3cm}
\sffamily
\begin{tabular}{m{4.5cm} p{13.5cm} }

 & \noindent\LARGE{\textbf{Knockout driven fragmentation of porphyrins$^\dag$}} \\
\vspace{0.3cm} & \vspace{0.3cm} \\

 & \noindent\large{Linda~Giacomozzi,$^{\ast}$\textit{$^{a}$}  Michael~Gatchell\textit{$^{a}$}, Nathalie~de~Ruette\textit{$^{a}$}, Michael~Wolf\textit{$^{a}$}, Giovanna~D'Angelo\textit{$^{a}$}\textit{$^{b}$}\textit{$^{c}$}, Henning~T. ~Schmidt\textit{$^{a}$}, 
Henrik~Cederquist\textit{$^{a}$} and Henning~Zettergren\textit{$^{a}$}} \\

\\

& \noindent\normalsize{We have studied collisions between tetraphenylporphyrin cations and He or Ne at center--of--mass energies in the 50~--~110 eV range. The experimental results were interpreted in view of Density Functional Theory calculations of dissociation energies and classical Molecular Dynamics simulations of how the molecules respond to He/Ne impact. We demonstrate that prompt atom knockout strongly contributes to the total destruction cross sections. Such impulse driven processes typically yield highly reactive fragments and are expected to be important for collisions with any molecular system in this collision energy range, but have earlier been very difficult to isolate for biomolecules.} \\

\end{tabular}

 \end{@twocolumnfalse} \vspace{0.6cm}

 ]

\renewcommand*\rmdefault{bch}\normalfont\upshape
\rmfamily
\section*{}
\vspace{-1cm}


\footnotetext{\textit{$^{a}$~Department of Physics, Stockholm University, Stockholm, SE-106 91, Sweden, E-mail: linda.giacomozzi@fysik.su.se}}
\footnotetext{\textit{$^{b}$~Faculdade de Ci\^{e}ncias, Universidade do Porto, 4169-007 Porto, Portugal}}
\footnotetext{\textit{$^{c}$~Departamento de Qu\'imica, M\'odulo 13, Universidad Aut\'onoma de Madrid, 28049 Madrid, Spain}}
\footnotetext{\dag~Electronic Supplementary Information (ESI) available: [details of any supplementary information available should be included here]. See DOI: 10.1039/b000000x/}



\section{Introduction}
\begin{figure}[h]
\centering
\includegraphics[width=.5\textwidth]{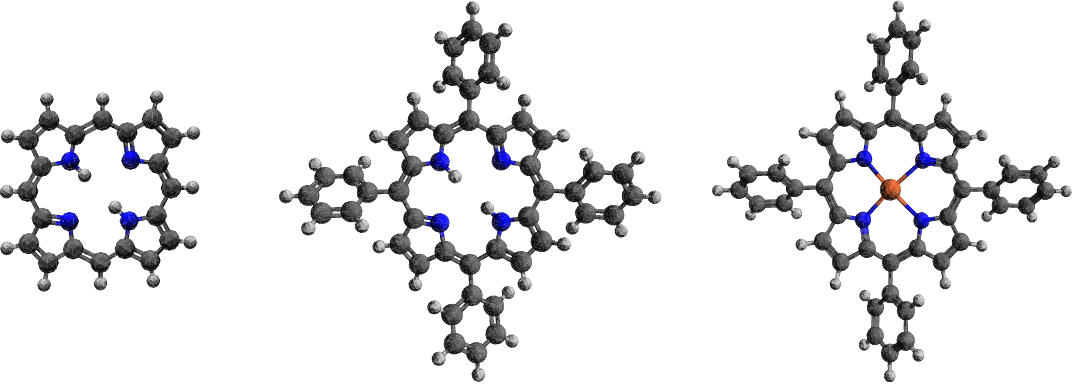}
\caption{Examples of porphyrin molecular structures. Left: Porphin (the simplest porphyrin). Middle: tetraphenylporphyrin (TPP). Right: metaltetraphenylporphyrin (MTPP). Carbons atoms are shown in black, H atoms in white, N atoms in blue and Fe in orange.}
\label{fgr:TPP_MTPP}
 \end{figure}

\indent When energy is transferred to a molecule, \emph{e.g.}\ in a collision with an ion or by the absorption of a photon, this energy will normally be redistributed over all its internal degrees of freedom within picoseconds. 
Such processes then lead to statistical (or thermally driven) fragmentation processes on longer timescales. Collisions with heavy particles may in addition lead to the knockout of individual atoms on femtosecond timescales in Rutherford-like heavy particle scattering processes. Since this fragmentation occurs before the energy has time to redistribute, it is often referred to as a non-statistical process.
Molecular fragmentation resulting from the knockout of single carbon atoms have recently been studied in experiments with Polycyclic Aromatic Hydrocarbons (PAHs),\cite{PAH_Stockett_2, PAH_Gatchell,PAH_review_Gatchell} hydrogenated PAHs,\cite{PAH_Gatchell_2, PAH_Wolf} Polycyclic Aromatic Nitrogen Heterocycles (PANH),\cite{PAH_PANH_Stockett} and fullerenes.\cite{PAH_Gatchell} 
From these experiments it has been concluded that fragments from knockout are much more reactive than intact molecules or fragments resulting from statistical decay processes.\cite{PAH_review_Gatchell,PAHs_Delaunay,PAH_Zettergren} However, to the best of our knowledge, direct evidence of knockout driven fragmentation has so far not been considered in connection with collision induced dissociation of biomolecules. One reason for this is that many biomolecules and their fragments have low dissociation energies and often fragment in several steps following the initial interaction. Another reason may be that biomolecules often have irregular three dimensional structures such that secondary knockout processes by the scattered projectile and/or by the atom that was initially knocked out make the interpretation of the data very involved. The porphyrins, however, have rigid structures resulting from delocalised $\pi$~orbitals (see Fig.~\ref{fgr:TPP_MTPP}). These aspects make them suitable for studies of non-statistical (knockout driven) fragmentation. \\
\indent Porphin is the simplest porphyrin and consists of four 5-membered nitrogen-containing carbon rings connected by carbon bridges (left structure in Fig.~\ref{fgr:TPP_MTPP}). Tetraphenylporphyrin (TPP) consists of a porphine with four phenyl groups (C$_6$H$_5$) attached to the meso-positions (middle structure, Fig.~\ref{fgr:TPP_MTPP}). Substituted porphines have different biological functionalities. Iron containing porphyrines (hemes) are for instance fundamental for oxygen transport in hemoproteins in blood cells\cite{Perutz135,Huynh1974} and magnesium porphyrines are important for the photosynthesis.\cite{porphyrin_book} Metaltetraphenylporphyrins (MTPP) are formed by substituting the central protons with a metal ion in a tetraphenylporphyrin molecule (M=Fe in the right structure in Fig. ~\ref{fgr:TPP_MTPP}). \\
\indent Several studies on MTPPs have been performed in order to investigate their stabilities under different forms of energetic processing, such as electron-impact (EI),\cite{Porphyrin_Feil, Porphyrin_Adler, Porphyrin_Beato, Porphyrin_Laycock,Meot-Ner} collisions with ions or atoms\cite{Porphyrin_Gozet, Porphyrin_Hayes,Porphyrin_Bernigaud} or photoinduced dissociation (PID).\cite{Porphyrin_Nuwaysir, Porphyrin_Castoro} This type of experiments may contribute to a better understanding of damage to biomolecules \textit{e.g.}\ in heavy ion therapy. Previous studies suggest that ions in the 10 -- 100 eV kinetic energy range may induce severe damage in biological tissue.\cite{Porphyrin_Deng} Such ions may be produced as secondary particles along high energy primary beams~\cite{Porphyrin_Deng2, Porphyrin_Deng, Porphyrin_Dang} in particular at penetration depths just before the primary beam is stopped (at the tail of the Bragg peak).\cite{Porphyrin_Bernigaud}\\
\indent Here we report a combined experimental and theoretical study of the fragmentation of three different tetraphenylporphyrin cations following collisions with He or Ne at center--of--mass energies in the 50 -- 110 eV range. In this collision energy regime, the energy transfer is mainly due to Rutherford--like scattering processes (nuclear stopping).\cite{PAH_review_Gatchell,PAH_Chen,PAH_Stockett_2,PAH_Stockett}\\

\section{Experimental technique}

\indent The experiments  were carried out with an ElectroSpray Ion (ESI) source at the single pass collision set--up at the DESIREE\cite{DESIREE1,DESIREE2} facility at Stockholm University, Sweden. The TPP sample (5,10,15,20-tetraphenylporphin) was dissolved in a 20~$\mu$M solution of methanol:acetic acid~(99:1), the FeTPP (5,10,15,20-tetraphenyl-21H,23H-porphinato iron(III)) chloride salt and the ZnTPP (3,10,15,20-tetraphenylporphyrinato zinc (II)) in a 20~$\mu$M solution of methanol:toluene~(50:50). These solutions were injected in the ESI source to bring the (M)TPP cations into the gas phase. The so formed (TPP+H)$^+$, FeTTP$^+$, and ZnTPP$^+$/(ZnTPP+H)$^+$ ions were then collected by a radio frequency ion funnel, mass selected by means of a quadrupole mass filter, and finally accelerated to kinetic energies in the 3.5~--~13~keV range. For the latter ion, it was not clear if the ion beam was in the protonated or the radical cationic form, but this is not expected to be important as the present results on non-statistical knockout of individual C or N atoms are similar for porphyrins with three different charge carriers. After the acceleration stage, the ion beam passed through a 40~mm long gas cell, containing He or Ne. We selected ion-beam energies such that we obtained center--of--mass energies, E$_{CoM}$, in the range of 50~--~80~eV and 110~eV for collisions with He, and Ne, respectively. The pressure in the gas cell was monitored by a capacitance manometer. The fragment ions from the collisions were focused by a set of cylinder lenses, energy--to--charge analysed by two pairs of electrostatic deflectors, and recorded with a position sensitive micro-channel plate detector.  The fragments have almost the same velocity as the parent ion before the collision. We may thus convert the kinetic energy per charge that we measure for the fragments to an (approximate) mass--to--charge scale. With this set--up, we are thus able to measure the fragment mass spectra, the absolute total destruction cross sections by measuring the attenuation of the primary beam as a function of the pressure in the gas cell and, thus, the absolute fragmentation cross sections for different fragmentation channels.\cite{PAH_Stockett} \\

\section{Calculation and simulation details}

\indent In order to investigate the importance of prompt nuclear scattering (knockout) processes, we have performed classical molecular dynamic (MD) simulations of collisions between TPP and He. We did not include electronic excitation processes here as they only give small contributions (few eV) in the present collision energy range.\cite{PAH_review_Gatchell} In the simulations, we have used the reactive Tersoff potential\cite{PAH_PANH_Stockett,Tersoff1,Tersoff2} to describe the breaking (and formation) of bonds in the TPP molecules. The interactions between the neutral projectile and all atoms in the molecule were computed using the Ziegler-Biersack-Littmark (ZBL) potential.\cite{ZBL} Previous studies on PAHs and related molecules using this simulation method have shown that experimental knockout cross sections are approximately a factor of 4/3 times larger than the simulated cross sections with He in the present energy range.\cite{PAH_PANH_Stockett,PAH_Stockett} Based on this, the simulated fragmentation cross sections for TPP~+~He collisions presented here have been scaled by a factor of 4/3.\\
\indent We have used Density Functional Theory (DFT) to calculate adiabatic dissociation energies for some of the dominant fragmentation channels of TPP and FeTPP molecules at the B3LYP/CC-pVDZ level of theory using the Gaussian 09 software package.\cite{Gaussian09} We have also calculated vibrational frequencies in order to ensure that actual minima on the potential energy surfaces have been located and we have determined the zero-point energy corrections.\\

\section{Results and discussion}
\begin{figure}[h!]
\centering
 \includegraphics[width=.5\textwidth]{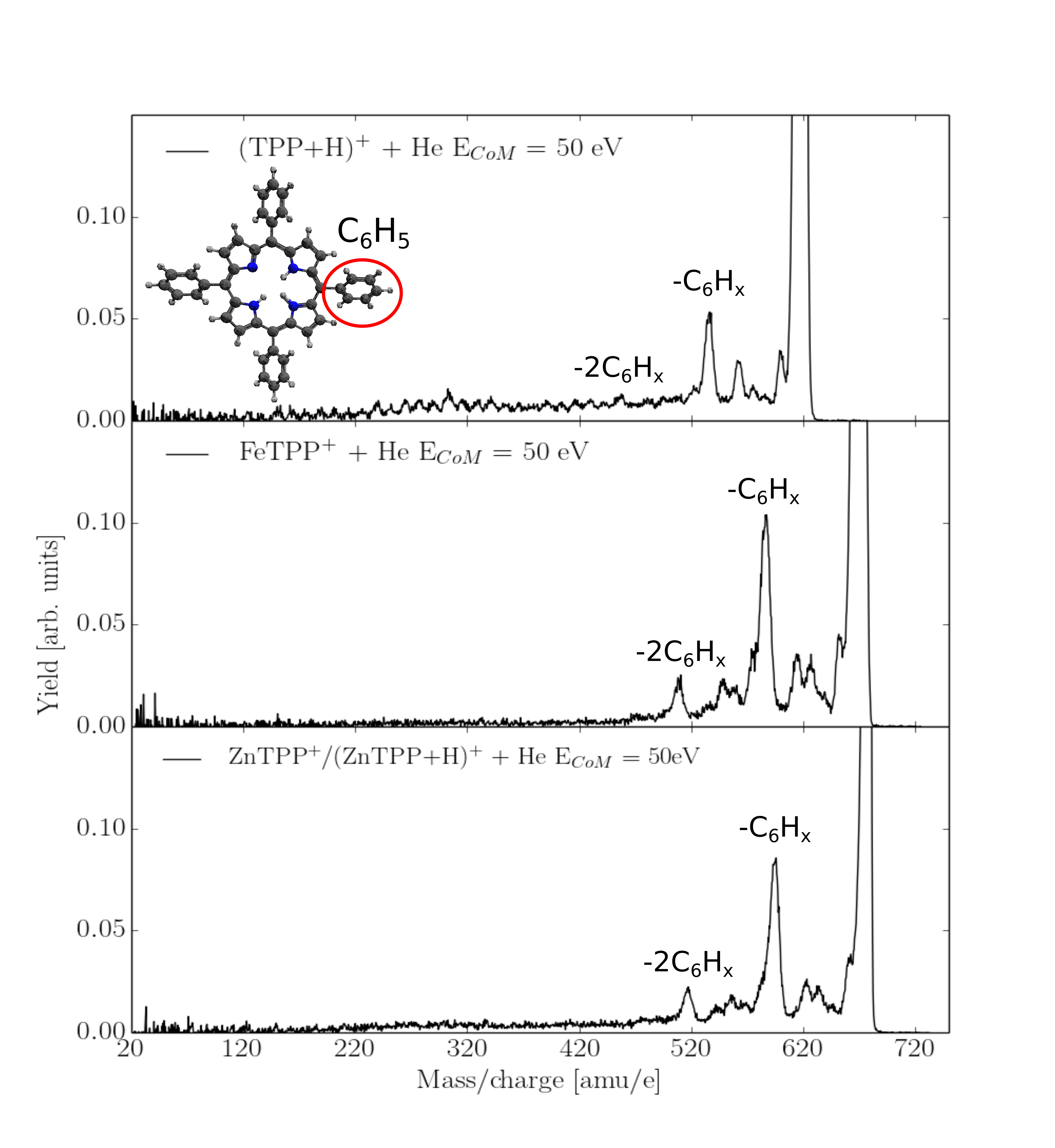}
 \caption{Collision induced dissociation spectra for (TPP+H)$^+$~+~He (top panel), FeTPP$^+$~+~He (middle panel) and ZnTPP$^+$/(ZnTPP+H)$^+$~+~He (bottom panel) collisions at 50~eV center--of--mass energy. The peaks corresponding to losses of one or two phenyl groups from the projectile cation are indicated as -C$_6$H$_x$ and -2C$_6$H$_x$, respectively.}
 \label{fig:CID_MTPP_TPP_He_50eV}
 \end{figure}

\begin{figure}[h!]
\centering
 \includegraphics[width=.5\textwidth]{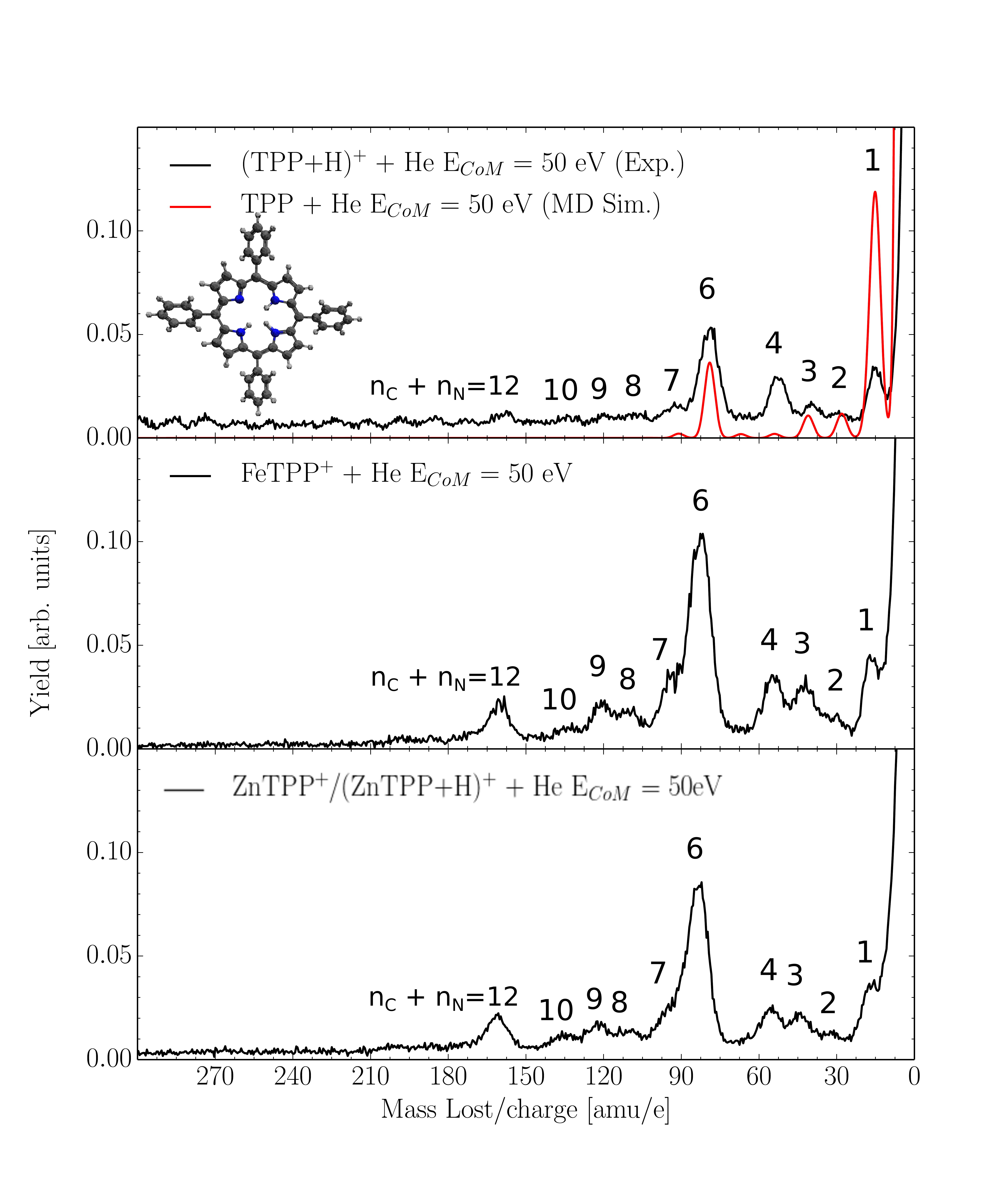}
 \caption{Zoom-ins of the mass spectra for (TPP+H)$^+$~+~He (in black) (top panel), FeTPP$^+$~+~He (middle panel) and ZnTPP$^+$/(ZnTPP+H)$^+$~+~He (bottom panel) collisions at 50~eV center--of--mass energy. The labels indicate the number of carbon (n$_C$) and nitrogen (n$_N$) atoms lost from the projectile ion. MD simulations for TPP + He are shown in red (top panel).}
 \label{fig:CID_MTPP_TPP_He_50eV_lost}
 \end{figure}

\begin{figure}[h!]
\centering
 \includegraphics[width=.5\textwidth]{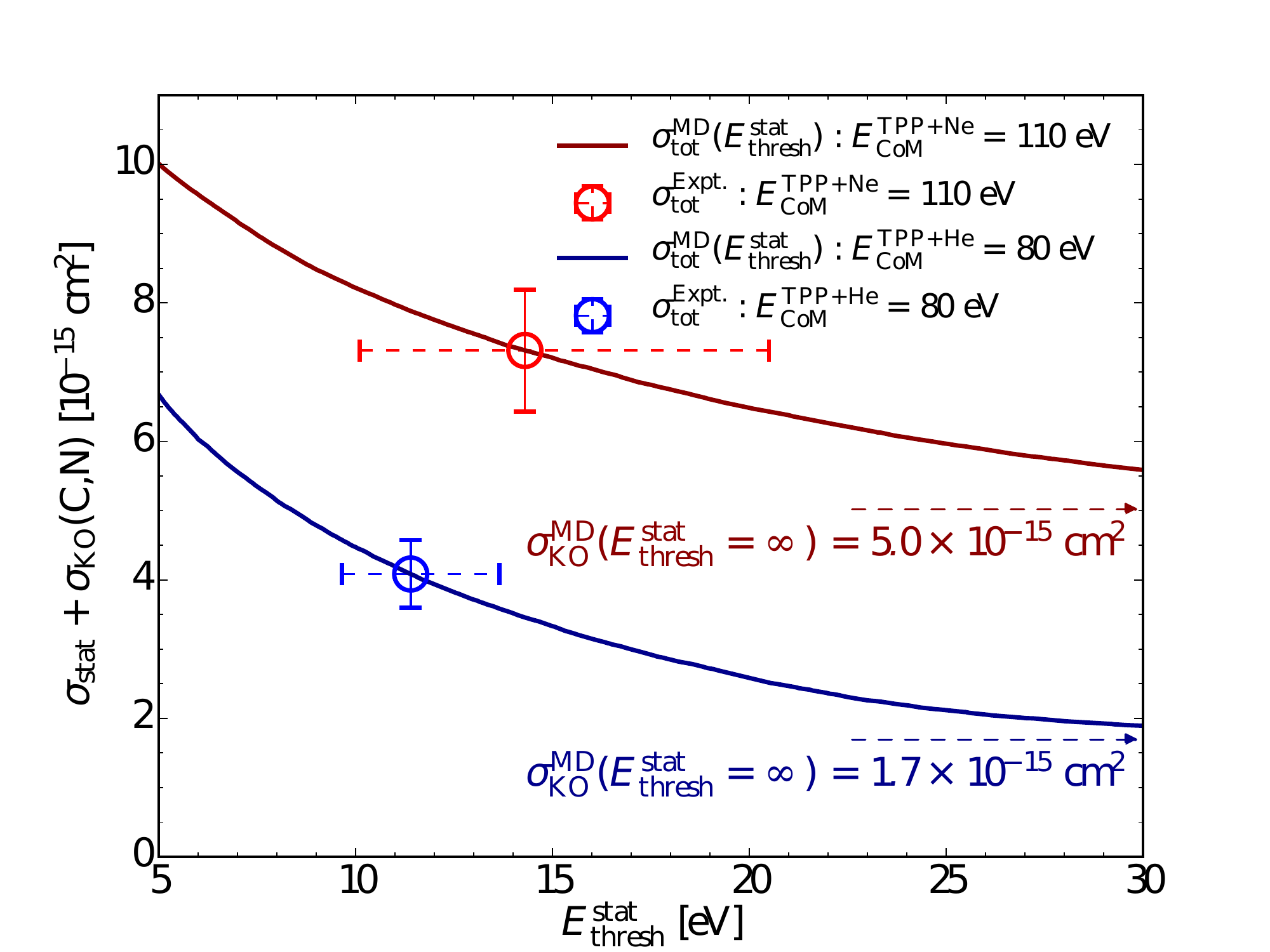}
 \caption{The calculated absolute total fragmentation cross sections (excluding H-loss) for TPP in collisions with He at 80~eV  (blue solid line) and Ne at 110~eV center--of--mass energy (red solid line) as functions of the assumed value of the threshold energy for statistical fragmentation. The circles are the experimental absolute fragmentation cross sections. The solid error bars represent the systematic uncertainties in the measured fragmentation cross section, which have been translated to uncertainties in the threshold energy (dashed error bars). The asymptotic values in the inset ($\sigma_{KO}^{MD}$) are total knockout cross sections from the MD simulations (dashed lines), excluding H-loss, which are about 40\% and 70\% of the absolute total cross section for He and Ne, respectively, at the present experimental conditions. }
 \label{fig:MD_Exp_StatisticalProcessees}
 \end{figure}

\begin{figure}[h!]
\centering
 \includegraphics[width=.5\textwidth]{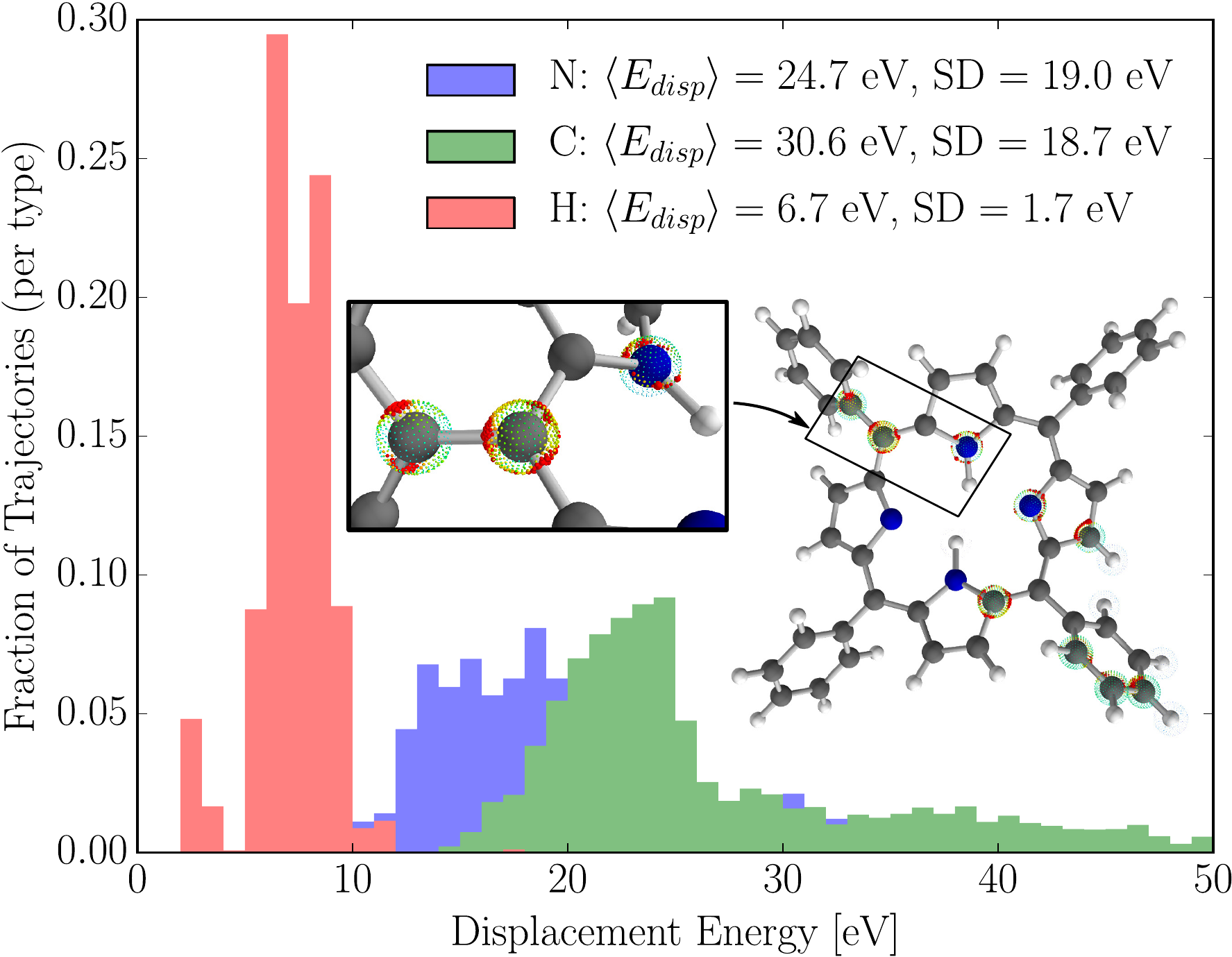}
 \caption{Distributions of displacement energies for prompt knockout of a single N (in blue), C (in green) or H (in red) atom from TPP molecules from the present classical MD simulations using the Tersoff force field.\cite{PAH_PANH_Stockett,Tersoff1,Tersoff2} The reported values correspond to the mean values of the displacement energies for each type of atom in the molecule (SD is the standard deviation of each distribution). The pictures represents an example of the displacement energies for two different carbons and one nitrogen position in the TPP molecule. Red dots in the inset show directions with the higher displacement energies and blue dots shows those with the lower displacement energies.}
\label{fig:disp_energ}
 \end{figure}

\indent In Fig.~\ref{fig:CID_MTPP_TPP_He_50eV}, we show mass spectra due to (TPP+H)$^+$~+~He~(top panel), FeTPP$^+$~+~He~(middle panel) and ZnTPP$^+$/(ZnTPP+H)$^+$~+~He~(bottom panel) collision at E$_{CoM}$~=~50 eV. 
The fragmentation yields in each panel are normalised to their respective total fragmentation cross section. The latter have been measured separately by means of the beam attenuation method. All three spectra display similar features with the dominant peak (off scale in Fig.~\ref{fig:CID_MTPP_TPP_He_50eV}) corresponding to the intact molecule and the second most prominent peak corresponding to the loss of a phenyl group (C$_6$H$_5$-loss indicated by the red circle in Fig.~\ref{fig:CID_MTPP_TPP_He_50eV}) which most likely is accompanied by the loss of an additional hydrogen atom.\cite{Meot-Ner} In the case of the MTPP molecules (M: metal, here Fe or Zn, middle and bottom spectra in Fig.~\ref{fig:CID_MTPP_TPP_He_50eV}), the absence of small fragments (with masses lower than 500 amu/e) show that the molecule is more stable with than without the metal. Other important features are the peaks between the primary and the single phenyl-loss peaks. In Fig.~\ref{fig:CID_MTPP_TPP_He_50eV_lost}, we show zoom-ins on this region. The peaks are labelled by the number of heavy atoms (C and/or N) that are lost from the parent ion. These peaks are separated by masses corresponding to a single C or N atom, typically accompanied by the loss of a small number of H atoms. \\
\indent The mass spectra in Fig.~\ref{fig:CID_MTPP_TPP_He_50eV_lost} are similar to those measured earlier for collisions between (TPP +H)$^+$ and He at 278~eV center--of--mass energy~\cite{Nielsen_porphyrin_1} and for collisions with protonated protoporphyrin~IX and Fe(III)-heme.\cite{Bernigaud_porphyrin_2,Kocks_porphyrin} There,\cite{Nielsen_porphyrin_1,Bernigaud_porphyrin_2,Kocks_porphyrin}  peaks corresponding to the loss of one or several heavy atoms were clearly observed but their origins were not discussed. The fragment distributions in Fig.~\ref{fig:CID_MTPP_TPP_He_50eV_lost} and those reported in Refs~\cite{Nielsen_porphyrin_1,Bernigaud_porphyrin_2,Kocks_porphyrin} are markedly different from those observed in other experiments.\cite{Porphyrin_Gozet,Porphyrin_Bernigaud,Porphyrin_Feil} Fragments from the loss of the phenyl groups were seen following 30 keV O$^{3+}$ and 70 eV electron impact on \emph{neutral} FeTPPCl.\cite{Porphyrin_Bernigaud,Porphyrin_Feil} In the latter case, fragmentation due to loss of 1--4 heavy atom were also observed but only with intensities that are orders of magnitude lower than in the present experiments. Gozet~\emph{et~al.}\ \cite{Porphyrin_Gozet} studied fragmentation of FeTPP$^+$ in low--energy multiple collisions with N$_2$. There, the most intense fragmentation peaks are due to the loss of one or two phenyl groups and loss of one or several H atoms, which correspond to the lowest energy dissociation channels according to our molecular structure calculations. For (TPP+H)$^+$, the dissociation energies are between 5.0~eV and 5.3~eV for loss of H atoms bonded to C atoms,  around 4~eV for H atoms connected to N atoms in the centre of the TPP ring and 4.8~eV for phenyl loss. In the case of FeTPP$^+$, dissociation energies range between 4.4~eV and 7.4~eV for H atoms bond to C atoms and the dissociation energy is 6.6~eV for phenyl loss (see the electronic supplementary information\dag). The losses of one or several heavy atoms (with or without hydrogen loss) are associated with significantly higher dissociation energies (except for the previously mentioned phenyl groups). Such fragmentation processes were not observed in the low energy collisions with N$_2$,\cite{Porphyrin_Gozet} but are clearly seen in the present mass spectra (see Fig.~\ref{fig:CID_MTPP_TPP_He_50eV_lost}). This suggests that the spectra recorded in Ref\cite{Porphyrin_Gozet} are dominated by statistical fragmentation processes, while we have significant contributions from knockout processes in the present work.\\
\indent In the top panel of Fig.~\ref{fig:CID_MTPP_TPP_He_50eV_lost}, we show a comparison between the experimental mass spectrum (in black) and classical MD simulation for direct atom knockouts (in red) in collision between protonated TPP and He at 50~eV center--of--mass energy. Because the TPP and MTPP experimental spectra display similar features, we may use the TPP MD results to guide the interpretation of the experimental results for all three porphyrins (see Fig.~\ref{fig:CID_MTPP_TPP_He_50eV_lost}). The experimental spectrum is normalised to the measured absolute total fragmentation cross section, while the simulated one is normalised to the absolute total cross section for prompt single and multiple-atom knockouts from the simulations. Note that neither of these cross sections include losses of H-atoms (see Table~\ref{tab:cross_section}). The experimental and simulated (heavy atom knockout) spectra partially display the same fragmentation channels, although the branching ratios are different. The simulations predict more fragments where one heavy atom has been lost and less fragments with four atoms lost than what is observed in the experiment. This may be explained by the difference in timescale between the MD simulations and the experiments. The simulations only consider prompt fragmentation by nuclear stopping processes and following secondary fragmentation on picosecond timescale. The experimental timescale is much longer (hundreds of microseconds) such that delayed statistical fragmentation and secondary fragmentation processes following knockout may have time to occur to much larger extents. The experimental results suggest that there is a preference for the loss of three heavy atoms following single atom knockouts. \\
\begin{table}[h]
\small
  \caption{\ Experimental absolute total fragmentation cross sections for 80~eV (TPP+H)$^+$~+~He and 110~eV (TPP+H)$^+$~+~Ne collisions (in units of 10$^{-15}$ cm$^2$) and MD simulations of the corresponding knockout (KO) cross sections. The experimental results ($\sigma$ $^{Exp}_{TOT}$) do not include the cross sections for single and multiple H-loss. The MD results are separated for knockout of heavy atoms ($\sigma_{KO}^{MD}$(C,N)) and for single and multiple H-atoms ($\sigma_{KO}^{MD}$(H)). }
  \label{tab:cross_section}
  \begin{tabular*}{0.5\textwidth}{@{\extracolsep{\fill}}lllll}
    \hline
    Target & E$_{CM}$ [eV] & $\sigma$ $^{Exp}_{TOT}$ & $\sigma$ $^{MD}_{KO}$(C,N) & $\sigma$$_{KO}^{MD}$(H) \\
    \hline
    He & 80  & 4.1 $\pm$ 0.5 & 1.69 $\pm$ 0.03 & 2.72  $\pm$ 0.04\\
 Ne & 110 & 7.3 $\pm$ 0.9 & 5.02 $\pm$ 0.05 &  2.19 $\pm$ 0.03\\
    \hline
  \end{tabular*}
\end{table}
\indent By comparing the simulated cross sections for depositing different amounts of energy in the TPP molecules and the measured total fragmentation cross section, we may determine a semi-empirical threshold energy for statistical fragmentation. The curves in Fig.~\ref{fig:MD_Exp_StatisticalProcessees} show the sums of calculated statistical and knockout cross sections as a function of the value of an assumed threshold energy for statistical fragmentation, E$^{stat}_{thresh}$. The data points are the present experimental values for the absolute total fragmentation cross sections (see Table 1).  Based on the experimental uncertainties and the statistical uncertainty of the simulations, E$^{stat}_{thresh}$ is $11.4 (\substack{+2.2\\-1.8})$ eV for collisions with He, and $14.3(\substack{+6.2\\-4.2})$ eV for collisions with Ne. A weighted average of E$_{thresh}^{stat}$~=~(12.8 $\pm$ 1.4) eV  gives the energy required to induce statistical fragmentation on the experimental (microsecond) timescale. This threshold energy is significantly higher than the dissociation energies, which is similar to what has been observed for PAHs\cite{PAH_statistical_anthracene} and fullerenes.\cite{PAH_statistical_fullerene} This reflects large heat capacities for these systems that protect them from prompt decay even when the excitation energy is larger than the dissociation energy.\\
\indent The asymptotic value of the calculated absolute total fragmentation cross section for infinite E$^{stat}_{thresh}$ corresponds to the total knockout cross section, $\sigma^{MD}_{KO}$(C,N) (dashed lines in Fig. \ref{fig:MD_Exp_StatisticalProcessees}) for the MD simulations. For He atoms colliding with TPP at 80 eV $\sigma^{MD}_{KO}$(C,N)~=~1.7~x~10$^{-15}$~cm$^{2}$, and for Ne $\sigma^{MD}_{KO}$(C,N)~=~5.0~x~10$^{-15}$~cm$^{2}$ for 110 eV. Relying on these numbers, we find that $\sim$40$\%$ ($\sim$70$\%$) of the total fragmentation cross section of the TPP in collision with He (for Ne) is due to non-statistical, \emph{i.e.}\ knockout,  fragmentation. These numbers are comparable to those for PAHs under similar conditions.\cite{PAH_Stockett_2}  \\ 
\indent The values of the knockout cross sections are governed by the energies required to permanently displace (remove) an individual atom from the molecule, which depend on the atom type and its chemical environment within the molecule. Here, we determine this molecular property (displacement energy) through MD simulations without using a specific projectile trajectory but displacing the atom in randomly generated straight line directions. In Fig. \ref{fig:disp_energ} we show distributions of displacement energies for prompt knockout of a single N (in blue) or C (in green) or H (in red) atom. The mean value of these broad energy distributions are 24.7~eV for knockout a N atom, 30.6~eV in the case of C and 6.7~eV for a H atom.
For the PAH molecule coronene (C$_{24}$H$_{12}$) the mean value of the displacement energy calculated with the same method for removing a C atom along any trajectory is slightly higher (35.4~eV)\cite{PAH_review_Gatchell} than in the case of porphyrin molecules . This and the similarities in the dissociation energies\cite{Holm} for M(TPP) and PAH cations, explains why the branching ratios between statistical and non-statistical fragmentation (knockout fragmentation) are comparable for the PAHs and the (M)TPP systems.\\

\section{Summary and Conclusions}

\indent In this paper, we have studied the importance of prompt atom knockout in collisions between tetraphenylporphyrin ions and noble gas atoms at 50 -- 110 eV center--of--mass energies. Through comparisons with the results from classical molecular dynamics simulations, we find that about 40 \% and 70 \% of the total fragmentation cross sections are due to such processes in collisions with He and Ne, respectively. This is similar to what has been reported for PAHs,\cite{PAH_review_Gatchell} which we attribute to comparable dissociation energies (about 5 eV) and displacement energies (about 30 eV) for PAHs, MTPPs, and TPPs.  \\
\indent Recent studies of PAHs and fullerenes have shown that atom knockout typically produce much more reactive fragments than thermally driven (statistical) fragmentation processes.\cite{PAHs_Delaunay,PAH_review_Gatchell} We plan to further investigate such processes in collisions with biomolecular ions, and the effect of embedding such molecules in a surrounding (cluster or solvent) environment. The latter may lead to efficient molecular growth processes, which have recently been observed following keV ion impact on clusters of fullerenes~\cite{PAH_review_Gatchell} and PAHs.\cite{PAHs_Delaunay}

\section{Acknowledgements}
\indent This work was supported by the Swedish Research Council~(Contract Nos.~2016-04181, 621-2015- 04990, and 621-2014-4501). We acknowledge the COST action CM1204 XUV/X-ray Light and Fast Ions for Ultrafast Chemistry~(XLIC) and the European Joint Doctorate on Theoretical Chemistry and Computational Modelling~(INT-EJD-TCCM).





\bibliography{rsc} 
\bibliographystyle{rsc} 

\end{document}